# Thickness identification of thin InSe by optical microscopy methods

*Qinghua Zhao, Sergio Puebla, Wenliang Zhang, Tao Wang\*, Riccardo Frisenda\*, Andres Castellanos-Gomez\**

Q. Zhao, Prof. T. Wang
State Key Laboratory of Solidification Processing, Northwestern Polytechnical University, Xi'an, 710072, P. R. China
Key Laboratory of Radiation Detection Materials and Devices, Ministry of Industry and Information Technology, Xi'an, 710072, P. R. China
E-mail: taowang@nwpu.edu.cn

Q. Zhao, S. Puebla, W. Zhang, Dr. R. Frisenda, Dr. A. Castellanos-Gomez
Materials Science Factory. Instituto de Ciencia de Materiales de Madrid (ICMM-CSIC), Madrid, E-28049, Spain.

E-mail: riccardo.frisenda@csic.es; andres.castellanos@csic.es

**Abstract:** Indium selenide (InSe), as a novel van der Waals layered semiconductor, has attracted a large research interest thanks to its excellent optical and electrical properties in the ultra-thin limit. Here, we discuss four different optical methods to quantitatively identify the thickness of thin InSe flakes on various substrates, such as $SiO_2$/Si or transparent polymeric substrates. In the case of thin InSe deposited on a transparent substrate, the transmittance of the flake in the blue region of the visible spectrum can be used to estimate the thickness. For InSe supported by $SiO_2$/Si, the thickness of the flakes can be estimated either by assessing their apparent colors or accurately analyzed using a Fresnel-law based fitting model of the optical contrast spectra. Finally, we also studied the thickness dependency of the InSe photoluminescence emission energy, which provides an additional tool to estimate the InSe thickness and it works both for InSe deposited on $SiO_2$/Si and on a transparent polymeric substrate.

## Introduction

Two-dimensional (2D) materials, thanks to their remarkable mechanical, electrical, optical, and optoelectronic properties, have been considered as promising candidates for future nano-electronic and optoelectronic device applications.[1-6] One of the interesting aspects of 2D materials is to investigate their electronic structure change with the reduction in thickness due to quantum confinement effects.[7-9] For example, monolayer $MX_2$ (M = Mo, W; X = S, Se) are expected to have a strong exciton binding energy of 0.5-1.0 eV in the single-layer limit and they show an indirect-to-direct band gap transition from bulk to single layer regime.[10-13] Black phosphorus (bP) is another example of 2D material with strong quantum confinement effects, its band gap changes from 0.2 eV in bulk to 1.5-2.0 eV for single-layer.[9, 14, 15] Therefore, the isolation of a new family of 2D materials by mechanical exfoliation is usually followed by experimental efforts to provide appropriate methods to determine the thickness of the exfoliated flakes. Until now, many optical techniques, such as transmittance/reflectance spectroscopy,[16, 17] photoluminescence,[10, 18] optical absorption,[19] Raman spectroscopy,[20] second harmonic generation as well as optical contrast,[21-25] have been employed as fast and non-destructive methods to determine the number of layers of different 2D materials.[7] Indium selenide (InSe), as a novel 2D semiconductor, recently has drawn a great research interests because of its interesting non-linear optical properties,[26, 27] superior electrical and optoelectronic device performance,[28-31] as well as the distinct mechanical features.[32, 33] Typically in literature, atomic force microscopy (AFM) has been the most used method to determine the thickness of 2D InSe flakes.[28, 34, 35] However, one of the limitations of the AFM technique is that it is a time-consuming measurement and thus it is not suitable for a rapid measurement of the thickness of 2D flakes over large areas.[36] Furthermore, the different interactions of the AFM tip with the flake and substrate can lead to large thickness discrepancy when testing ultrathin flakes and introduce artifacts in the measurement.[34, 37-41]

In this context, a comprehensive comparison of different optical microscopy based methods [24, 28, 34, 35, 42-46] to determine the thickness of InSe flakes would be very helpful for the development of the research on this novel 2D material. That is precisely the goal of this manuscript, we summarize four different optical microscopy methods to estimate the thickness of InSe flakes deposited on different substrate. The employed methods include techniques partially reported in previous reports (analysis of the optical contrast and photoluminescence (PL) spectroscopy) [24, 28, 34, 35, 42-46]  as well as the development of new optical microscopy based techniques, not demonstrated for InSe, like the quantitative analysis of transmission mode optical microscopy images and the apparent color of the flakes using a quantitative color chart. We believe that gathering this information together is helpful to promote the research on this novel 2D semiconductor.

**Results and discussions**

- Optical transmittance

Thin InSe flakes are fabricated using mechanical exfoliation method from single crystal bulk InSe grown by the Bridgman method.[47] The InSe bulk crystal was firstly cleaved with Nitto tape (SPV 224) and then was transferred onto a Gel-film (WF 4 x 6.0 mil Gel-Film from Gel-Pak, Hayward, CA, USA) substrate. **Figure 1a** shows the transmission mode optical images of two InSe flake that have been deposited onto the Gel-film substrate recorded using an optical microscope equipped with a color CMOS digital camera. In the transmission mode images the regions covered with InSe flakes are darker as the InSe flakes absorb part of the incoming light. The intensity on the InSe flakes depends monotonically on their thickness. Actually, the optical images are composed of three separate color channels which give information of the light transmitted in the red, green and blue regions of electromagnetic spectrum. The transmittance ($T$) at a given position of the image can be calculated, for each channel, by dividing the intensity by the average intensity of the bare substrate. **Figure 1b** shows the line profiles of 1-$T$ versus the positions (indicated by the solid lines with arrows in panel a), extracted from the blue channel intensity of the images. The line profiles reach lowest values (0 percent) in the bare Gel-film regions. In the flake regions, the values of 1-$T$ tend to increase upon the thickness growth of the thin InSe. The 1-$T$ value allows us to quantitatively determine the thickness of thin InSe flakes. Optical pictures of tens of different InSe flakes were recorded and then the transmittance value (taken from the blue channel images) of each flake was extracted. In order to measure their thickness, all the InSe flakes were subsequently deposited onto 270 nm $SiO_2$/Si substrates with a dry-transfer method, and an AFM was employed to determine their thicknesses in dynamic mode. We present the data collected from 18 different thickness regions in thin InSe flakes used to correlate the 1-$T$ values to their height measured with AFM in **Figure 1c** (the top x-axis indicates the InSe thickness). From the plot it is evident that the data, in the thickness range experimentally probed, follow a linear trend and we use linear regression to fit them. We find a slope of 1.4 ± 0.1 %/nm and an offset of around 5 nm of flake thickness. Note that when testing an ultrathin flake with AFM, the offset between 2D flakes and substrate can be as large as several nanometers, which can be explained by the different interactions of the AFM tip with the InSe flake and $SiO_2$/Si substrate as well as the adsorbates on the InSe surface and interfaces. [37-40, 48] To further demonstrate the easiness of using this calibration to determine the thickness of thin InSe, we marked the number of layers of the investigated InSe flake in **Figure 1a** in left axis of panel b, where we assume a value 0.8 nm for a InSe single layer. It is worth to note that this method also can be used for the thickness determination of freely-suspended InSe samples.

- Apparent color

It is well known that graphene, transition metal dichalcogenides (TMDCs), $TiS_3$, franckeite, mica, antimonene and $MoO_3$ flakes deposited on a $SiO_2$/Si substrate presents different colors depending on their thickness and on the $SiO_2$ thickness.[16, 49-54] A comprehensive analysis of the apparent colors can yield a quick guide to estimate the thickness of 2D materials in a similar fashion as the thickness of $SiO_2$ capping layers on Si wafers is estimated from its interference color. **Figure 2a** shows the atomic force microscopy (AFM) topography of seven InSe flakes with thicknesses ranging from ∼ 4 nm to ∼ 90 nm, and **Figure 2b** shows their corresponding optical microscopy images on 270 nm $SiO_2$/Si substrate (the $SiO_2$ thickness has been experimentally determined by reflectometry with ± 0.5 nm uncertainty). By comparing the AFM thickness and the optical image colors of these InSe flakes, one can build up a color-chart correlating the apparent color of the InSe deposited on the top of 270 nm $SiO_2$/Si substrate with their corresponding thickness. Since the interference colors of the thin InSe flakes have a strong dependence on the underlying substrate thickness, it is necessary to use the substrates with specific $SiO_2$ thickness (in our case we provide a thickness color chart that will be valid for 270 nm $SiO_2$/Si substrates).

- Optical contrast analysis

The number of layers of InSe flakes that have been deposited on the $SiO_2$/Si substrate can be further determined more accurately by quantitatively analyzing their reflection spectra. In **Figure 3a**, we show a schematic drawing of our four-media air/InSe/270 nm $SiO_2$/Si optical system. The optical contrast (C) of the InSe flake on the $SiO_2$/Si substrate can be calculated from the spectra measured on the bare substrate ($I_{sub}$) and the one measured on the InSe flake ($I_{InSe}$):

$$C = \frac{I_{InSe} - I_{sub}}{I_{InSe} + I_{sub}}. \qquad (1)$$

The differential reflectance spectra are acquired in normal incidence with a modified metallurgical microscope (BA 310 MET-T, Motic) and experimental operation details have been described in our previous work.[17]

The optical contrast of this kind of multilayer system can be also modelled with high accuracy using a Fresnel law-based model that takes into account the light reflected and transmitted at the different interfaces.[55] The reflection coefficient in a Fresnel model with four media can be expressed as:[56]

$$r_{InSe} = \frac{r_{01}e^{i(\Phi_1+\Phi_2)}+r_{12}e^{-i(\Phi_1-\Phi_2)}+r_{23}e^{-i(\Phi_1+\Phi_2)}+r_{01}r_{12}r_{23}e^{i(\Phi_1-\Phi_2)}}{e^{i(\Phi_1+\Phi_2)}+r_{01}r_{12}e^{-i(\Phi_1-\Phi_2)}+r_{01}r_{23}e^{-i(\Phi_1+\Phi_2)}+r_{12}r_{23}e^{i(\Phi_1-\Phi_2)}}, \quad (2)$$

where the subscript 0 refers to air, 1 to InSe, 2 to SiO$_2$ and 3 to Si. Under normal incident condition, $\Phi_i = 2\pi\tilde{n}_i d_i/\lambda$ is the phase shift induced by the propagation of the light beam in the media *i*, in which $\tilde{n}_i$, $d_i$ and $\lambda$ are the complex refractive index, thickness of the media and wavelength, respectively; $r_{ij} = (\tilde{n}_i - \tilde{n}_j)/(\tilde{n}_i + \tilde{n}_j)$ is the Fresnel coefficient at the interface between the media *i* and *j*.

The reflection Fresnel coefficient in a three media (the case of the bare substrate without being covered by InSe flake) is expressed as:

$$r_{sub} = \frac{r_{01}+r_{12}e^{-i2\Phi_1}}{1+r_{01}r_{12}e^{-i2\Phi_1}} \quad (3)$$

Where sub index 0 is air, 1 is SiO$_2$ and 2 is Si. Using equations (2) and (3), we can calculate the optical contrast by firstly calculating the reflected intensity of both situations as

$$R_{InSe} = |\overline{r_{InSe}}r_{InSe}|, R_{sub} = |\overline{r_{sub}}r_{sub}|. \quad (4)$$

Then the optical contrast can be obtained through the following equation (5) that correlates the reflected intensity by the bare substrate ($R_{sub}$) with the reflected intensity by the InSe flake ($R_{InSe}$) as:

$$C = \frac{R_{InSe}-R_{sub}}{R_{InSe}+R_{sub}}. \quad (5)$$

Interestingly, we found that using the thickness of the SiO$_2$ layer (determined by reflectometry), the thickness of the InSe flakes (measured with AFM), the reported refractive indexes for air, SiO$_2$ and Si, and assuming a thickness-independent refractive index $\tilde{n}_i = 2.7 - i0$ for the InSe flakes we can reproduce the experimental optical contrast spectra accurately.[57-59] Moreover, by considering that the thickness of the flake is an unknown parameter we can determine the thickness of the InSe flakes by calculating optical contrast spectra for different InSe thicknesses and computing which thickness provides the best matching with the experimental data. **Figure 3b** shows how we compare the optical contrast experimentally measured in an InSe flake (9 nm thick according to the AFM measurement) with the modelled optical contrast assuming a thickness in the range of 5-13 nm. There is a clear best match for a thickness of 9 nm, illustrating the potential of this method to determine the thickness of InSe flakes. The inset in **Figure 3b** shows the square of the difference between the measured contrast and the calculated one as a function of the thickness assumed for the modelling. The plot shows a well-defined minimum centered at a thickness of 9 nm. In **Figure 3c** we further compared the experimental optical contrast curves obtained from other three InSe flakes with thicknesses of 14 nm, 21 nm and 34 nm (top panel) and the fit curves (bottom panel) with the Fresnel-based model. We find the fitted flake thicknesses of 15 nm, 24 nm and 32 nm are consistent with the flake heights measured by AFM, with a small error of ± 2 nm. In order to benchmark this thickness determination method for InSe, in **Figure 3d** we compare the flake height values determined with AFM from 16 flakes from 4 nm to 90 nm thick with the value obtained following the discussed optical contrast fit method. In this plot, one can find a slope value of 1.05 ± 0.02 marked by the straight line indicates the good agreement between the thicknesses of thin InSe flakes measured by AFM and the fit to the Fresnel law-based model (a perfect agreement in this representation would yield a linear trend of slope equal to 1). Finally, we argue that the choice for the thickness-independent refractive index is motivated by the negligible variation of the band structure of InSe with the thickness in the range probed experimentally (4-90 nm, see also Fig. 4b).

- Photoluminescence (PL)

Photoluminescence (PL) is another important method to identify the thickness of 2D semiconducting flakes. In contrast to $MX_2$, layered InSe exhibits a direct-to-indirect band gap transition while reducing the flake thickness from bulk to monolayer regime. The large quantum efficiency supported by direct band gap in few-layer InSe regime indicates the PL can be employed for layer number *n* determination of thin InSe in a larger thickness range at room temperature.[28, 34] **Figure 4a** shows the photoluminescence spectra that are collected from four InSe flakes with the thicknesses of 6 nm, 9 nm, 12 nm and 22 nm, respectively. With the increase of the flakes thickness, one can observe an obvious redshift of the maxima of the PL peak: from 1.34 eV (6 nm) to 1.27 eV (22 nm). Note that the PL peak emission energy from the spectra were determined using a Gaussian fit. More importantly, such well-shaped PL spectra can be collected both on the InSe flakes deposited on the transparent (*e.g.* polycarbonate, PC) and opaque (*e.g.* $SiO_2$/Si) substrates. In **Figure 4b**, we show the results of a statistical analysis of the PL emission peak as a function of InSe flake thickness extracted from 42 different InSe flakes deposited on PC and $SiO_2$/Si substrate. The thickness dependence of the exciton optical transition and hence the optical band gap of InSe flakes, as obtained from the PL-peak maxima, follows a trend that seems to be nicely described in terms of quantum-size confinement effects on the direct band gap of InSe. Thus we can follow this PL emission energy ($E_{PL}$) as a function of thickness of InSe flake (*h*, nm) using the equation based on the quantum well confinement effect,

$$E_{PL} = \frac{a}{h^2} + b, \tag{6}$$

which the black line indicates the best fit to the present data points with the parameters are determined to be *a* = 3.25 ± 0.44 eV · $nm^2$, *b* = 1.27 ± 0.01 eV. Our fit presents a good agreement with the results previously reported data in the literature.[28, 34, 35, 42-45]

| Method | Quant./Qual. | Formula ($h$ in nm) * | Relative Error |
|---|---|---|---|
| Transmission (blue channel) $(1 - T)$ | Quantitative | $h = a * (1 - T)$<br><br>$a = (71 \pm 7)$ nm | 10% ($h$ < 30 nm) |
| Optical contrast spectroscopy | Quantitative | / | ~5% ($h$ < 100 nm) |
| Photoluminescence peak energy $E_{PL}$ | Quantitative | $h = \sqrt{\dfrac{a}{E_{PL} - b}}$<br><br>$a = (3.25 \pm 0.44)$ eV·nm$^2$<br><br>$b = (1.27 \pm 0.01)$ eV | 10% ($h$ ~ 6 nm)<br><br>30% ($h$ ~ 10 nm)<br><br>60% ($h$ ~ 15 nm) |
| Apparent Color | Qualitative | / | / |

**Table 1**: **Summary of the optical identification methods proposed in this work.** * $h$ is the estimated InSe thickness in nm.

**Conclusion**

In summary, we have provided several fast and non-destructive complementary optical methods to evaluate the number of layer of thin InSe deposited on different substrates. The results from the four proposed techniques are summarized in Table 1. For InSe flakes deposited on transparent substrates (such as Polydimethylsiloxane, PDMS; polycarbonate, PC), we find the transmittance of InSe flake in the blue region of the electromagnetic spectrum decreases at a rate of 1.4 %/nm upon increasing the InSe thickness. We further demonstrate how to determine the flakes thickness of InSe that have been deposited on the Si substrate covered with a specific thickness of SiO$_2$ using optical contrast spectroscopy. On the one hand one can quick estimate the InSe flake thickness with an error of ± 3 nm directly from the observed apparent color of the flake based on a calibrated color versus thickness chart. On the other hand, one can first measure the optical contrast spectrum of a InSe flake under investigated and then fit the spectrum using a Fresnel law-based model to obtain its thickness information, which can reach a more accurate result with the lower uncertainty of ± 2 nm (in the case of 270 nm $SiO_2$/Si substrate). As the last yet powerful tool, we also calibrate the photoluminescence emission peak energy of InSe flakes (both deposited on PC and $SiO_2$/Si substrates) as a function of their thicknesses in the range of 5 layers to 30 layers. We believe that these calibrations based on above optical methods also can be easily reproduced for the other 2D semiconductors.

**Materials and Methods**

**Sample fabrication.** Thin InSe flakes were fabricated with mechanical exfoliation method with Nitto tape (Nitto Denko® SPV 224) and then onto a Gel-film (WF 4 x 6.0 mil Gel-Film from Gel-Pak, Hayward, CA, USA). After inspection with optical microscope (Motic® BA310 MET-T), the images of the selected InSe flakes were recorded under transmission mode with a microscope digital camera (AmScope, MU1803), which are employed for blue channel transmittance analysis. Finally, the flakes were dry-transferred onto 270 nm $SiO_2$/Si or polycarbonate (PC) substrates for AFM, optical contrast, and photoluminescence measurements.

**Optical contrast characterization.** The optical contrast of the flake on 270 nm $SiO_2$/Si substrates was measured with a micro-reflectance setup that has been described in our previous work.[17] A fiber-coupled Thorlabs spectrometer was connected to a Motic BAMET 310 metallurgical microscope equipped with both transmission and epi-illumination halogen lamps. The reflection of the sample used for the characterization of spectral contrast between the transferred InSe flakes and $SiO_2$/Si substrates.

**AFM measurements.** The thickness of thin InSe flakes was measured using two commercial AFM setups: (1) an ezAFM (by Nanomagnetics) atomic force microscope operated in dynamic mode. The cantilever used is Tap190Al-G by BudgetSensors with force constant 40 Nm$^{-1}$ and resonance frequency 300 kHz. (2) A Nanotec AFM system that has been operated in contact mode with a NextTip (NT-SS-II) cantilever.

**ACKNOWLEDGEMENTS**

This project has received funding from the European Research Council (ERC) under the European Union's Horizon 2020 research and innovation programme (grant agreement n° 755655, ERC-StG 2017 project 2D-TOPSENSE). EU Graphene Flagship funding (Grant Graphene Core 3, 881603) is acknowledged. Funding from the Spanish Ministry of Economy, Industry and Competitiveness through the grant MAT2017-87134-C2-2-R is acknowledged. RF acknowledges support from the Netherlands Organization for Scientific Research (NWO) through the research program Rubicon with project number 680-50-1515 and from the Spanish Ministry of Economy, Industry and Competitiveness through a Juan de la Cierva-formación fellowship (2017 FJCI-2017-32919). SP acknowledges the fellowship PRE2018-084818. QHZ acknowledges the grant from China Scholarship Council (CSC) under No. 201706290035. T.W. acknowledges support from the National Key R&D Program of China: No. 2016YFB0402405. This work is also sponsored by Innovation Foundation for Doctor Dissertation of Northwestern Polytechnical University.

**COMPETING INTERESTS**

The authors declare no competing financial interests.

**FUNDING**

Spanish Ministry of Economy, Industry and Competitiveness: Juan de la Cierva-formación fellowship 2017 FJCI-2017-32919

EU H2020 European Research Council (ERC): ERC-StG 2017 755655

This is the authors' version (post-peer reviewed) of the following article:
Q. Zhao et al., "*Thickness identification of thin InSe by optical microscopy methods*"
Advanced Photonics Research (2020)
https://doi.org/10.1002/adpr.202000025

EU Graphene Flagship: Grant Graphene Core 3, 881603

National Natural Science Foundation of China: 51672216

**FIGURES**

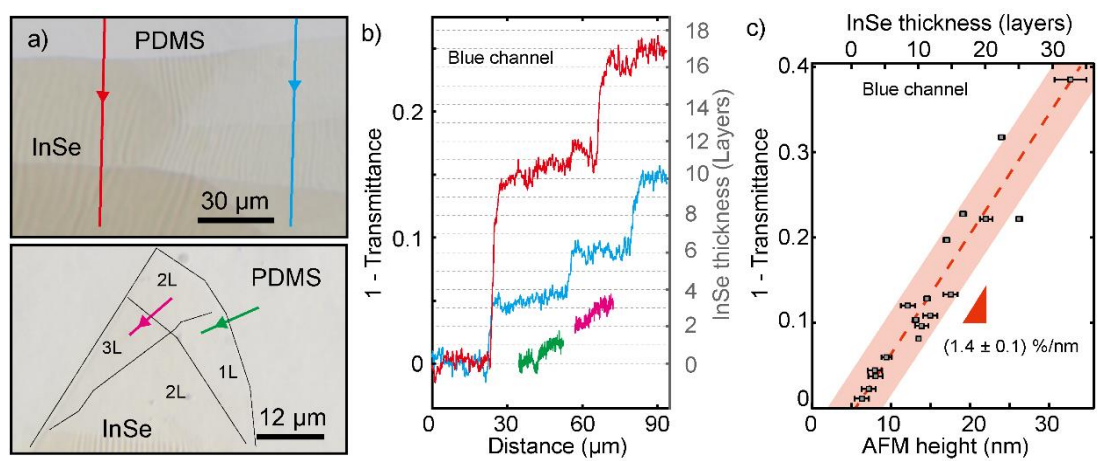

**Figure 1**: **Optical transmission of thin InSe.** a) Optical image of two mechanically exfoliated thin InSe flakes on the Gel-film recorded with an optical microscopy working under transmission mode. The colored lines in the image correspond to the intensity

profiles in panel (b). b) The normalized blue channel intensity line profiles of 1-transmittance recorded as a function of InSe flake regions with various thicknesses indicated upon the arrows in (a). Note that the bare Gel-film takes a unity transmittance. c) The relationship between InSe flake heights determined by atomic force microscopy (AFM) versus the 1-transmittance.

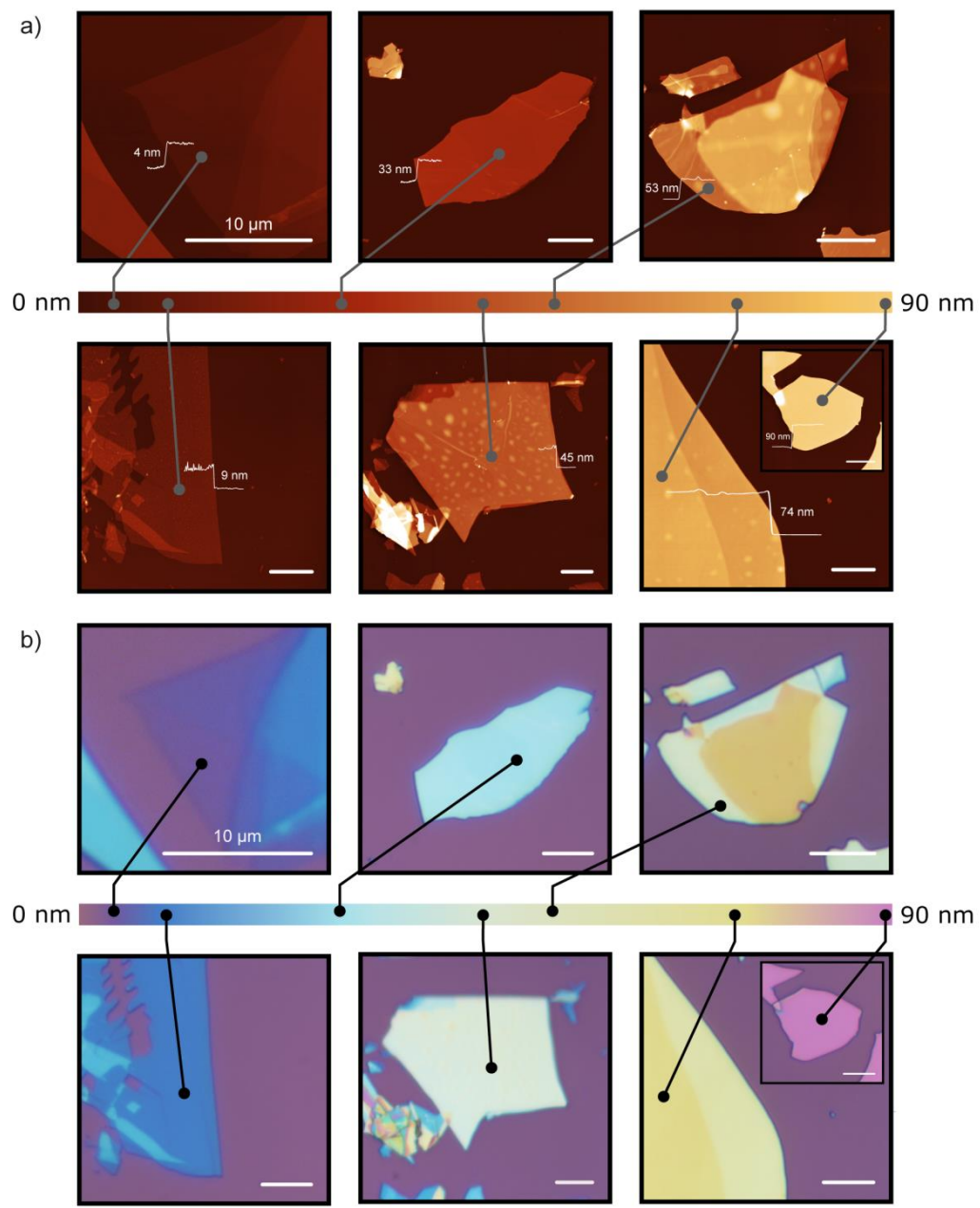

**Figure 2: Apparent colors of thin InSe upon flake thicknesses on 270 nm SiO$_2$/Si substrates.** a) Atomic force microscopy (AFM) characterizations of the exfoliated thin InSe flakes with different thickness deposited on 270 nm SiO$_2$/Si substrates. b) Optical pictures of the thin InSe flakes shown in panel (a) and the color bar indicates the apparent color dependency of thin InSe upon the flake thickness up to 90 nm. All the scale bars are 10 μm.

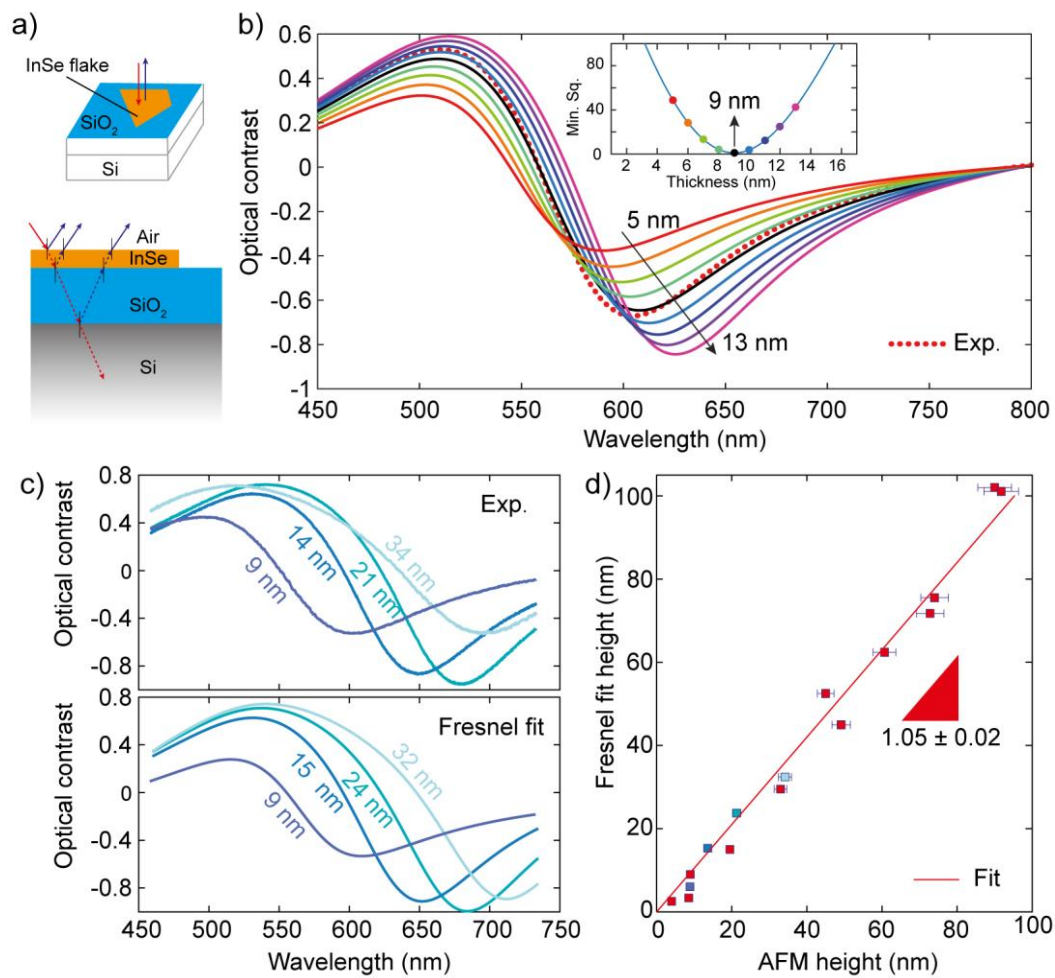

**Figure 3: Optical contrast of thin InSe on 270 nm SiO$_2$/Si substrates.** a) Schematic drawing of the four-media system (air/InSe/SiO$_2$/Si) from which the optical contrast spectra calculated. b) Comparison between experimental optical contrast spectrum of a 9

nm thick InSe flake (red dot line) and the calculated ones (colored solid lines), with the thickness ranging from 5 nm to 13 nm, and deposited on 270 nm SiO$_2$/Si substrates. The inset shows the minimum square value as a function of flake thickness. c) Experimental optical contrast spectra of thin InSe flakes with different thickness on 270 nm SiO$_2$/Si substrates (top panel) and the corresponding Fresnel fits (bottom panel). d) Comparison between the thickness determined with AFM measurements and Fresnel fit. The experimental data represented by the red squares, the error bars demonstrates the uncertainty of the AFM measurements, and the red line guides a fit with a slope of 1.05 ± 0.02.

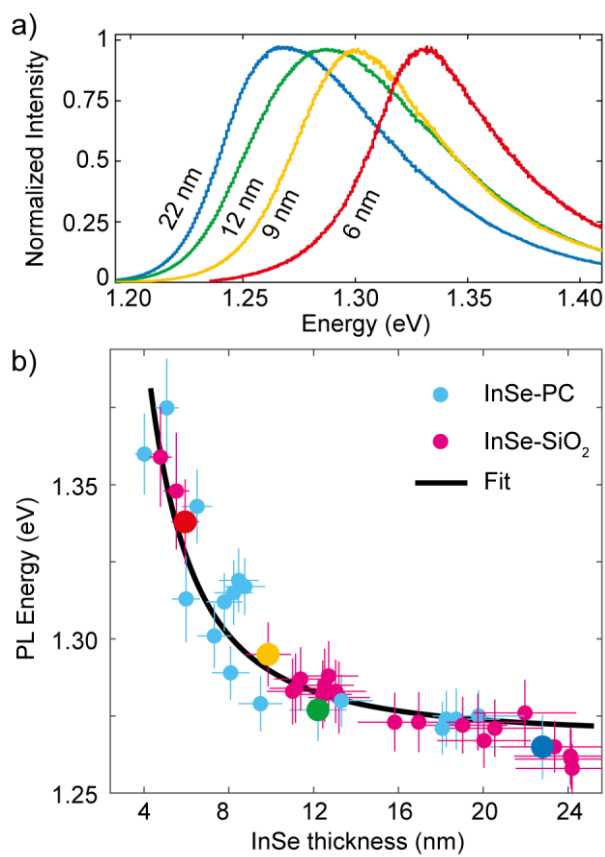

**Figure 4**: **Photoluminescence (PL) vs. InSe thicknesses.** a) Photoluminescence spectra of thin InSe flakes with different number of layers. b) Emission energy of the PL spectra

as a function of InSe flake thickness extracted from 42 different InSe flakes deposited on PC (red dots) and $SiO_2$/Si (blue dots) substrates. The error bars come from the uncertainty of the thickness determination and extraction of the PL emission energy. The black line indicates a best fit based on quantum well confinement effect.